\documentclass[conference]{IEEEtran}
\IEEEoverridecommandlockouts
\usepackage{cite}
\usepackage{amsmath,amssymb,amsfonts}
\usepackage{algorithmic}
\usepackage{graphicx}
\usepackage{textcomp}
\usepackage{xcolor}
\usepackage{multirow}
\usepackage[super]{nth}
\def\BibTeX{{\rm B\kern-.05em{\sc i\kern-.025em b}\kern-.08em
    T\kern-.1667em\lower.7ex\hbox{E}\kern-.125emX}}
\begin{document}

\title{Siamese Sleep Transformer For Robust Sleep Stage Scoring With Self-knowledge Distillation and Selective Batch Sampling
\footnote{
\thanks{20xx IEEE. Personal use of this material is permitted. Permission from IEEE must be obtained for all other uses, in any current or future media, including reprinting/republishing this material for advertising or promotional purposes, creating new collective works, for resale or redistribution to servers or lists, or reuse of any copyrighted component of this work in other works.}
}
\thanks{This work was partly supported by Institute for Information $\&$ Communications Technology Promotion (IITP) grants funded by the Korea government(MSIT) (No. 2015-0-00185: Development of Intelligent Pattern Recognition Softwares for Ambulatory Brain-Computer Interface, No. 2017-0-00451: Development of BCI based Brain and Cognitive Computing Technology for Recognizing User’s Intentions using Deep Learning, No. 2019-0-00079: Artiﬁcial Intelligence Graduate School Program, Korea University, and No. 2021-0-02068: Artificial Intelligence Innovation Hub).}\\ 
}

\author{\IEEEauthorblockN{Heon-Gyu Kwak}
\IEEEauthorblockA{\textit{Dept. Artificial Intelligence} \\
\textit{Korea University} \\
Seoul, Republic of Korea \\
hg\_kwak@korea.ac.kr} \\

\and

\IEEEauthorblockN{Young-Seok Kweon}
\IEEEauthorblockA{\textit{Dept. Brain and Cognitive Engineering} \\
\textit{Korea University} \\
Seoul, Republic of Korea \\
youngseokkweon@korea.ac.kr} \\
\and

\IEEEauthorblockN{Gi-Hwan Shin}
\IEEEauthorblockA{\textit{Dept. Brain and Cognitive Engineering} \\
\textit{Korea University} \\
Seoul, Republic of Korea \\
gh\_shin@korea.ac.kr} \\
 \and
%
}

\maketitle 

\begin{abstract}
In this paper, we propose a Siamese sleep transformer (SST) that effectively extracts features from single-channel raw electroencephalogram signals for robust sleep stage scoring. Despite the significant advances in sleep stage scoring in the last few years, most of them mainly focused on the increment of model performance. However, other problems still exist: the bias of labels in datasets and the instability of model performance by repetitive training. To alleviate these problems, we propose the SST, a novel sleep stage scoring model with a selective batch sampling strategy and self-knowledge distillation. To evaluate how robust the model was to the bias of labels, we used different datasets for training and testing: the sleep heart health study and the Sleep-EDF datasets. In this condition, the SST showed competitive performance in sleep stage scoring. In addition, we demonstrated the effectiveness of the selective batch sampling strategy with a reduction of the standard deviation of performance by repetitive training. These results could show that SST extracted effective learning features against the bias of labels in datasets, and the selective batch sampling strategy worked for the model robustness in training.
\end{abstract}

\begin{small}
\end{small}

\section{INTRODUCTION}
Sleep stage scoring is one of the major tasks in the sleep research field, widely used for the diagnosis of sleep disorders and assessment of sleep quality \cite{sleepdpc, hjkim}. Usually, sleep stage scoring is manually done by human sleep experts via reading polysomnograms (PSG). The PSG is the set of bio-signals epoched in 30-seconds that contains an electroencephalogram (EEG), electrocardiogram, electrooculography, etc. Sleep experts score every 30-second PSG epoch as wake (W), non-rapid eye movement (NREM) stage 1 (N1), NREM stage 2 (N2), NREM stage 3 (N3), NREM stage 4 (N4), and rapid eye movement (REM) according to Rechtschaffen $\&$ Kales (R$\&$K), but N3 and N4 can be treated as the same stage according to the American Academy of Sleep Medicine (AASM) manuals \cite{interrater}. This manual scoring task requires high labor intensity and need of expertise with huge time consumption. For this reason, the requirement for automatic sleep stage scoring research has gradually increased. 

Traditionally, most of the brain-computer interface (BCI) research used machine learning techniques based on statistical methods, which extract and select features from data manually based on the professionalized knowledge of the researchers \cite{won2017motion, zhang2017hybrid, thung2018conversion, lee2018high, lee2019connectivity}. These approaches showed promising results for using the machine learning methods in the BCI research, but have the limitation that the performance could be highly dependent on the knowledge of the researcher. To address this limitation, deep learning techniques, which are one branch of machine learning, have been proposed to extract and select important features from data by the model itself, in the various fields of BCI \cite{kim2019subject, kwon2019subject, jeong2020brain}. Likewise in the field of automatic sleep stage classification, the most of research also leverages deep learning techniques nowadays, and successfully increased the performance \cite{deepsleepnet,attnsleep,xsleepnet}. 

However, their performances are specific to a certain dataset and would not be constant across various datasets. Sleep stage can be scored differently by human experts because of their differences in experience and thought \cite{interrater, cesari2021interrater}. This would result in the bias of labels in datasets and causes the inconstancy of the model performance. This inconstancy could be caused by another reason, which is from the instability of deep learning methods or a model. Depending on the initial direction of training and the noise of data, the model makes different predictions for every trial and this causes the instability of the performance \cite{zheng2016improving}. These problems prevented applying automatic sleep stage scoring models in a real-world environment even though they have shown significant advances in the last few years.

To overcome these problems, we propose a novel deep learning architecture named Siamese Sleep Transformer (SST) with self-knowledge distillation and selective batch sampling. To alleviate the bias of labels in datasets, we focused on extracting effective features that can be observed across datasets if they are in the same labels. For this objective, we designed our model based on the Siamese network and transformer \cite{transformer, siam}. Class-wise self-knowledge distillation loss was used to regularize the class-wise differences between subjects and datasets \cite{skd}. Selective sampling methods are the strategy to store and reuse random samples when the model achieved good or bad performances trained with them, and we applied this strategy in our model to reduce the instability of repetitive training.

\begin{figure*}[ht]
    \centering
    \includegraphics[width=13cm]{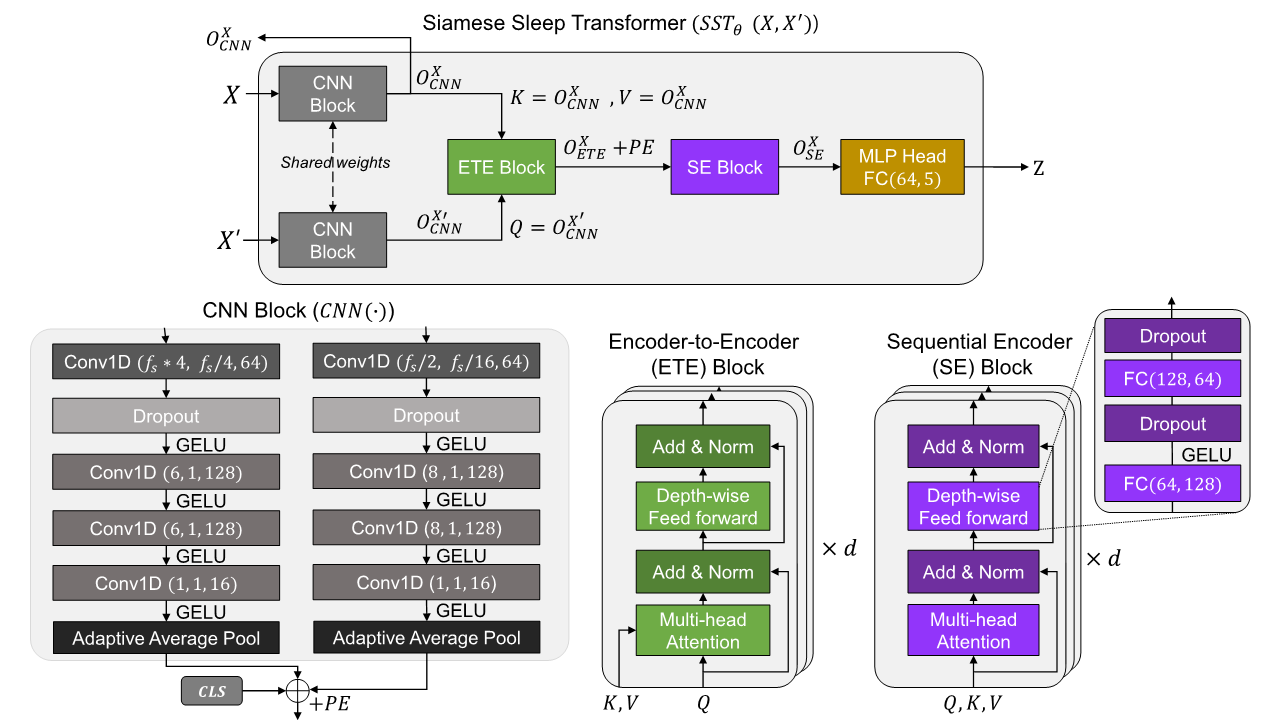}
    \caption{An architecture of Siamese sleep transformer (SST). $X$ and $X'$ are inputs of SST and $Z$ is the output of SST. $d$ is a number of encoder-to-encoder (ETE) and sequential encoder (SE) blocks. GELU is the Gaussian error linear unit. $PE$ is the positional encoding value. $CLS$ denotes the class token. $Q$, $K$, and $V$ are query, key, and value vectors, respectively.}
    \label{fig:sst}
\end{figure*}

\section{Proposed Method}
We hypothesized that there are general features between two data samples when they share the same label. The SST model was designed to extract those effective learning features to be robust to the bias of labels in the dataset. For this reason, SST takes two inputs $X \in R^{B\times S\times C \times T}$ and $X^\prime$, where $B$ is the batch size, $S$ is the length of a sequence, $C$ is the number of channels, and $T$ is the number of time points. $X^\prime$ is a randomly sampled data that has the same sequence of labels $Y\in R^{B\times S}$ with $X$. Note that $Y$ represents five classes. In the inference step, we set $X$ and $X^\prime$ to be the same. In this study, we set $C=1$ because we use single-channel EEG data as input, and set $S=20$ and $T=30f_s$, where $f_s$ is the sampling rate of the signal. 

\subsection{Siamese Sleep Transformer}
The SST model consists of two CNN blocks, an encoder-to-encoder (ETE) block, a sequential encoder (SE) block, and a multi-layer perceptron (MLP) head. Each encoder block can be stacked $d$ times like in the original transformer architecture \cite{transformer}. 

A CNN block $CNN(\cdot)$ consists of two CNN modules with different kernel-sized layers and an adaptive average pooling layer. We set kernel sizes for CNN layers as $4f_s$ and ${f_s}/{2}$ according to previous works \cite{deepsleepnet,attnsleep},  to extract both frequency and temporal features of EEG signals effectively. 

Each CNN module takes input sequence $X$ as input. To extract the feature with 1D-convolutional layers, the input data is reshaped into $(B \times S, C, T)$, and passes through CNN modules. After two CNN modules extract features from $X$, they are pooled by adaptive average pooling layer to fit their number of dimensions for multi-head attention layers, and then concatenated into feature-wise ($\text{Concat}(\cdot)$) with a class token $T^{X}_{cls}$. The class token is used for better performance and reducing computational cost \cite{bert, vit}. The outputs of the CNN module $O^{X}_{CNN} \in R^{(B \times S, N+1, D)}$ are then added by positional encoding values $PE(\cdot)$ like the following equation: 

\begin{align}
        PE(N) =& \sin \left({pos \over 10000^{2[N/2] \over D}} - {{(1+(-1)^{N}) \pi \over 4}} \right), \\
        O^{X}_{CNN} =& \text{Concat} \left(T^{X}_{cls}, CNN(X)) \right)+PE
\end{align}
where $N$ is the number of features, $pos$ is the position of features, and $D$ is the dimension of features. $X^{\prime}$ follows the same procedure with $X$ by the second CNN block that shares weight with the first CNN block for $X$ and returns as $O^{X\prime}_{CNN}$. Since our goal is the extraction of general features, we used the cosine similarity function as a loss function to make $O^{X}_{CNN}$ and $O^{X\prime}_{CNN}$ similar \cite{siam, dylee}. 

The ETE block designed for the model attends to class-wise general features so that reduces within-class variation from different subjects. With multi-head attention layers in the ETE block, the model can attend to features that have high similarity between $O^{X}_{CNN}$ and $O^{X\prime}_{CNN}$. The ETE block consists of multi-head attention layers, add $\&$ norm layers, and depth-wise feed-forward layers. When input vectors $Q$ and $C$ were given, attention scores from the transformer is calculated by the following equation: 
\begin{align}
        Q_i =& QW^Q_i, K_i = CW^K_i, V_i = CW^V_i, \\
        H^{(C,Q)}_i =& \text{Softmax} \left({Q_i(K_i)^T \over {\sqrt{D}}} \right)V_i,\\
        M^{(C,Q)} =& \text{Concat} \left(H_1^{(C,Q)}, \dots, H_A^{(C,Q)} \right)W^{M}
\end{align} where $Q_i$,$K_i$,$V_i$ denote query, key, and value vectors respectively, and $A$ is the number of multi-head. In the multi-head attention layer, $O^{X}_{CNN}$ is used as $K$ and $V$ vectors and $O^{X\prime}_{CNN}$ is used as $Q$ vector. After the multi-head attention layer, add $\&$ norm layers and the depth-wise feed-forward network is calculated by the following equation:
\begin{align}
    L^{(C,Q)} =& \text{LayerNorm} (M^{(C,Q)}+Q ), \\
    D^{(C,Q)} =& GELU (L^{(C,Q)}W^{E1} )W^{E2}, \\
    E^{(C,Q)} =& \text{LayerNorm} (D^{(C,Q)}+L^{(C,Q)}) )
\end{align}
where $GELU$ is the Gaussian error linear unit and $W^{E}$ is the weight of the encoder $E$. The output of ETE block is then pooled into $T_{cls}^{X}$, so the output shape of ETE block changed from $O^X_{ETE} \in R^{(B\times S, N+1, D)}$ to $O^X_{ETE} \in R^{(B\times S, 1, D)}$ and reshaped into $(B,S,D)$ to be used for the input of SE block $\tilde{O}$. The SE block is designed to extract the sequential features between sleep stage transitions. It has the same structure as the ETE block, but we add positional encoding value and calculate attention scores between epochs of the out-of-ETE blocks. The output of the SE was then passed to an MLP head to classify the sleep stage (Fig. \ref{fig:sst}). Finally, SST produces $Z$ by the following equation:
\begin{align}
    O^X_{ETE} =& E^{ \left(O^{X^\prime}_{CNN},O^{X}_{CNN} \right)}, \\
    O^X_{SE} =& E^{ \left(\tilde{O},\tilde{O} \right)} \left(\tilde{O} = \text{Pooling} \left(O^X_{ETE} \right) + PE\right),\\
    Z =& ReLU(O^X_{SE})W^{MLP}.
\end{align}
\begin{figure}[t]
    \centering
    \includegraphics[width=7cm]{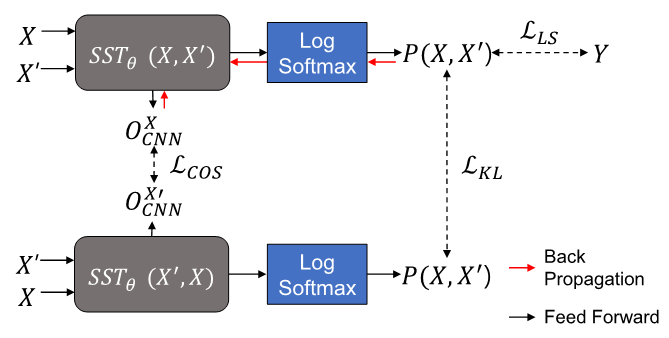}
    \caption{An overall training structure of Siamese Sleep Transformer. $\mathcal{L}_{COS}$ is a cosine similarity loss function. $\mathcal{L}_{LS}$ denotes label smoothing loss function. $\mathcal{L}_{KL}$ is a self-knowledge distillation loss function.}
    \label{fig:loss}
\end{figure}

\subsection{Loss Function for Training SST}
For inputs $(X,X^\prime)$ and a 5-dimensional one-hot target $Y$, the SST produces the logit vector $Z(X,X^\prime) = [z_1,...,z_5]$, and predicted probabilities $P(X,X^\prime) = [p_1,...,p_5]$ is calculated using label smoothing by following equation:
\begin{equation}
p_i \left((X,X^\prime);\tau) \right)=-\log \left({{\exp(z_i / \tau)}\over{\Sigma^5_{j=1}{exp(z_j / \tau)}}} \right)
\end{equation}
where $\tau$ denotes temperature scaling parameter for better distillation \cite{skd}. 
Sleep stage scoring has a high risk of mislabeling since they are determined by human experts. Also, sleep stage scoring can be differently done by each expert \cite{interrater}. To alleviate this problem, we used the label smoothing technique \cite{labelsmooting}. Label smoothing loss function $\mathcal{L}_{LS}$ is given by: 
\begin{align}
&\mathcal{L}_{LS} = \alpha \left(-\Sigma^5_{i=1}(p_i/B)) + (1-\alpha)(-\Sigma^5_{i=1}log(p_{i}Y_{i})\right) 
\end{align}
where $\alpha$ is a hyper-parameter that determines the smoothing level.

To extract features that have small within-class variation, SST compares outputs of CNN block using cosine similarity $sim_{cos}$ between $O^{X}_{CNN}$ and $O^{X^\prime}_{CNN}$. We also used class-wise self-knowledge distillation loss function for training, to regularize the difference between domains \cite{skd}, which adds KL divergence loss between $X$ and $X^\prime$.

\begin{align}
&\mathcal{L}_{COS} = 1-sim_{cos} \\
&\mathcal{L}_{KL}= KL\left(P(X,X^\prime)||P(X^\prime,X)\right) \\
&\mathcal{L}_{SST} = \mathcal{L}_{LS} + \mathcal{L}_{COS} + \lambda \cdot \tau^2 \cdot \mathcal{L}_{KL}
\end{align}
where $KL(\cdot)$ denotes the Kullback-Leibler divergence and $\lambda$ is the weight parameter. We set the temperature scaling parameter $\tau=5$ and the weight parameter $\lambda=1$ in this study (Fig. \ref{fig:loss}).

\subsection{Selective Batch Sampling} 
The instability of the model performance could be a problem for SST since $X^\prime$ is the sequence of randomly selected samples. $X^\prime$ can be either set of easy or difficult samples to learn extracting good features. To alleviate the training instability caused by randomly selected $X^\prime$, we propose selective batch sampling. We applied a sampling memory that stores the easiest and most difficult samples. The sampling memory stores $X^\prime$ if the model achieved the highest or the lowest validation loss so far. Hyper-parameter $p_0$ is set as the sampling probability. The easy and difficult sets of $X^\prime$ are sampled with $p_0$ probability. Other $X^\prime$ are sampled with $1-2p_0$ probability. Since there is a class imbalance problem in the sleep dataset, we also used balanced batch sampling to prevent the model over-fit to the dominant class labels \cite{utime}. 

\section{Experiments and Results}
\subsection{Dataset}
In this study, we used two different open datasets, Sleep-EDF-20 and sleep heart health study (SHHS) \cite{sleepedf, shhs} for training and evaluating our model. Sleep-EDF is the dataset that contains PSG and hypnogram of 78 healthy Caucasian subjects aged from 25 to 101 years old. We selected 20 subjects and pre-process the dataset following settings from previous works \cite{deepsleepnet, attnsleep, xsleepnet}. The SHHS is a large-scale dataset collected by various centers. In this study, we selected 329 healthy subjects and pre-process the dataset like previous work \cite{attnsleep}.

\subsection{Experimental Settings}
The training session consisted of 10,000 training steps in total. For every 100 training steps, we validated the model using a validation set, which is 10\% of randomly selected data from the training set. We also applied early stopping to prevent the overfitting of the model. When the model's validation performance was not increased for 10 continuous validation tasks the training session ended. The model that achieved the highest validation performance was used for the inference session.

The training optimizer was Adam \cite{adam, bhlee}, with weight decay of 0.0001, a learning rate of 0.001, and betas $(b_1,b_2)=(0.9,0.999)$. We also used gradient clipping with max norm value 5 to prevent the gradient exploding which can be caused by Gaussian error linear unit activation function \cite{gelu} in the model. The input sequential length for each batch was 20 and the batch size was 64. Consequently, 1,280 input data were used in a training step and 1,280 predicted labels were returned. For model hyperparameters, we set the encoder depth ($d$) as 3. The number of multi-headed attention heads was 8, and the number of dimensions for each head was 8. The number of dimensions for the feed-forward network in each encoder block was 128.

\subsection{Performance of Datset-to-Dataset Transfer Scoring} 
To identify the robustness of the model across two different datasets, we conducted transfer learning. For this experiment, we resampled the SHHS data from 125Hz to 100Hz, which is the same sampling rate as Sleep-EDF data, to keep the number of model parameters (especially CNN) when we transfer the model to another dataset. Experiments were conducted in two cases, one is training the model with the Sleep-EDF dataset and testing with SHHS dataset (EDF-to-SHHS), and another is the opposite case (SHHS-to-EDF). We did not conduct the fine-tuning session after the transfer, to focus on the evaluation of the capacity of extracting effective features of each model. 

We trained the DSN and ATS by following the proposed settings in their original papers \cite{deepsleepnet,attnsleep}. Each model was trained with 10,000 training steps with validation for every 100 training steps, and models with the highest validation score were used for the inference session. Table \ref{tab:sleepstage} showed the result of the experiment. In the case of EDF-to-SHHS transfer, SST could achieve the highest mean $\mathcal{F}1$-score among baselines. For the result of class-wise $\mathcal{F}1$-score, SST achieved the highest performance on the inference of W, N2, and N3 classes.

In the case of SHHS-to-EDF transfer, SST also achieved the highest mean $\mathcal{F}1$-score among baselines and could achieve higher performance on the inference of W, N1, and N3 classes. Especially, SST achieved remarkable performance on the inference of the N1 class, which has the lowest inter-rater reliability compared to the other classes \cite{interrater}.  

\begin{table}[t]
\caption{Transfer Performance of Sleep Stage Scoring.}
\label{tab:sleepstage}
\centering
\begin{tabular}{c|c|ccccc|c}
\hline
Data & Model & W          & N1            & N2            & N3            & R           & Mean          \\ \hline
EDF     & DSN   & 0.51          & 0.09          & 0.70          & 0.28          & \textbf{0.65} & 0.45          \\
to       & ATS   & 0.58          & \textbf{0.14} & 0.69          & 0.48          & 0.62          & 0.50          \\
SHHS      & SST   & \textbf{0.62} & 0.11          & \textbf{0.72} & \textbf{0.60} & 0.59          & \textbf{0.53} \\ \hline
SHHS      & DSN   & 0.66          & 0.00          & \textbf{0.73} & 0.62          & 0.65          & 0.53          \\
to       & ATS   & \textbf{0.82} & 0.02          & \textbf{0.73} & 0.67          & \textbf{0.71} & 0.59          \\
EDF     & SST   & \textbf{0.82} & \textbf{0.20} & 0.72          & \textbf{0.69} & 0.56          & \textbf{0.60} \\ \hline

\end{tabular}
\end{table}

\begin{table}[b]
\caption{Performance of SST in Different Loss Settings.}
\centering
\label{tab:loss}
\begin{tabular}{c|ccccc|c}
\hline
Loss   & W & N1   & N2   & N3   & R  & Mean         \\ \hline
only $\mathcal{L}_{CE}$     & 0.61 & 0.03  & 0.57 & 0.65 & 0.56 & 0.48         \\
only $\mathcal{L}_{LS}$     & 0.63 & 0.00  & 0.66 & \textbf{0.69} & 0.56 & 0.51         \\
\begin{tabular}[c]{@{}c@{}} $\mathcal{L}_{LS}+\mathcal{L}_{KL}$   \end{tabular} 
& 0.75 & 0.13 & 0.66 & 0.68 & \textbf{0.61} & 0.57         \\ 
\begin{tabular}[c]{@{}c@{}} $\mathcal{L}_{LS}+\mathcal{L}_{COS}$   \end{tabular}     
& 0.73 & 0.00 & 0.54 & 0.61 & 0.58 & 0.49 \\ 
$\mathcal{L}_{SST}$     & \textbf{0.82} & \textbf{0.20}  & \textbf{0.72} & \textbf{0.69} & 0.56 & \textbf{0.60}         \\ \hline
\end{tabular}
\end{table}

\subsection{Effects of Loss Function on Performance} 
We designed our model to extract effective features across different datasets by using $\mathcal{L}_{LS}$, $\mathcal{L}_{COS}$, and $\mathcal{L}_{KL}$ loss functions. To evaluate the effects of each loss function on our model, we conducted an ablation study on each loss function. We trained our model with different settings of loss functions for this experiment: only $\mathcal{L}_{CE}$, only $\mathcal{L}_{LS}$, $\mathcal{L}_{LS}$ and $\mathcal{L}_{KL}$, $\mathcal{L}_{LS}$ and $\mathcal{L}_{COS}$, and $\mathcal{L}_{SST}$. 

Table \ref{tab:loss} showed the result of the ablation study on loss functions. Note that we reported $\mathcal{F}_1$ score. Training with soft labeling (only $\mathcal{L}_{LS}$) showed higher performance than with hard labeling (only $\mathcal{L}_{CE}$). This result would indicate that bias of labels in the dataset exists, and soft labeling could deal with this problem. Training without $\mathcal{L}_{KL}$ or $\mathcal{L}_{COS}$ resulted in lower performance than using both of them in the training. It was considered that both class-wise regularization with $\mathcal{L}_{KL}$ and learning general features of a label $\mathcal{L}_{COS}$ could significantly affect the performance of dataset-to-dataset transfer. Especially, $\mathcal{L}_{KL}$ was more contributed to increment of performance than $\mathcal{L}_{COS}$. 

\begin{table}[t]
\caption{Statistics of Inference Performance in Different Settings of Selective Batch Sampling.}
\centering
\label{tab:selective}
\begin{tabular}{cccc}
\hline
Sampling           & $\mathcal{F}1$                                                               & Acc.                                                             & Kappa                                                            \\ \hline
No Selective       & \begin{tabular}[c]{@{}c@{}}0.750\\ ($\pm$0.060)\end{tabular}          & \begin{tabular}[c]{@{}c@{}}0.837\\ ($\pm$0.055)\end{tabular}          & \begin{tabular}[c]{@{}c@{}}0.785\\ ($\pm$0.072)\end{tabular}          \\
Easy and Difficult & \begin{tabular}[c]{@{}c@{}}0.747\\ (\textbf{$\pm$0.037})\end{tabular}          & \begin{tabular}[c]{@{}c@{}}0.832\\ ($\pm$0.039)\end{tabular}          & \begin{tabular}[c]{@{}c@{}}0.780\\ ($\pm$0.051)\end{tabular}          \\
Easy               & \begin{tabular}[c]{@{}c@{}} \textbf{0.759}\\ ($\pm$0.043)\end{tabular} & \textbf{\begin{tabular}[c]{@{}c@{}}0.844\\ ($\pm$0.038)\end{tabular}} & \textbf{\begin{tabular}[c]{@{}c@{}}0.795\\ ($\pm$0.049)\end{tabular}} \\ \hline
\end{tabular}%
\end{table}

\subsection{Effectiveness of Selective Sampling} 
To analyze the effectiveness of selective sampling on training stability, we trained our model in different conditions of selective samplings and compared standard deviations of repetitive training performances for each condition. The conditions were no selective sampling, selective sampling with the easy and difficult samples, and selective sampling with the easy samples. We trained the model using Sleep-EDF by 10 repetitive times with each condition. As shown in Table \ref{tab:selective}, the mean of inference performance increased at selective sampling with an easy sample compared to other conditions. Also, the standard deviation of inference performance decreased compared to other conditions. Training without selective sampling resulted in the highest standard deviation of repetitive training performances. This result showed selective sampling worked for the improved model performance and training stability.

\section{Conclusion}
In this paper, we proposed SST with selective batch sampling and the method to apply self-knowledge distillation loss for sleep stage scoring. Our proposition demonstrated competitive performance for dataset-to-dataset transfer classification and reduced the standard deviation of performance by repetitive training. These results showed our approach worked for the robustness of the model against label bias in datasets and the instability of deep learning techniques. We expect that the proposed methods can be improved when it is combined with research about domain-specific features like \cite{suk2014predicting} and \cite{jeong2020decoding}. Thus, Our future work would focus on leveraging this research on our methods.

\bibliographystyle{IEEEtran}
\bibliography{REFERENCE}

\begin{thebibliography}{10}
\providecommand{\url}[1]{#1}
\csname url@samestyle\endcsname
\providecommand{\newblock}{\relax}
\providecommand{\bibinfo}[2]{#2}
\providecommand{\BIBentrySTDinterwordspacing}{\spaceskip=0pt\relax}
\providecommand{\BIBentryALTinterwordstretchfactor}{4}
\providecommand{\BIBentryALTinterwordspacing}{\spaceskip=\fontdimen2\font plus
\BIBentryALTinterwordstretchfactor\fontdimen3\font minus
  \fontdimen4\font\relax}
\providecommand{\BIBforeignlanguage}[2]{{%
\expandafter\ifx\csname l@#1\endcsname\relax
\typeout{** WARNING: IEEEtran.bst: No hyphenation pattern has been}%
\typeout{** loaded for the language `#1'. Using the pattern for}%
\typeout{** the default language instead.}%
\else
\language=\csname l@#1\endcsname
\fi
#2}}
\providecommand{\BIBdecl}{\relax}
\BIBdecl

\bibitem{sleepdpc}
Q.~Xiao \emph{et~al.}, ``Self-supervised learning for sleep stage
  classification with predictive and discriminative contrastive coding,'' in
  \emph{Proc. IEEE Int. Conf. Acoust. Speech Signal Process. (ICASSP)}, 2021,
  pp. 1290--1294.

\bibitem{hjkim}
H.-J. Kim, M.~Lee, and S.-W. Lee, ``{E}nd-to-end automatic sleep stage
  classification using spectral-temporal sleep features,'' in \emph{Conf. Proc.
  IEEE Eng. Med. Biol. Soc. (EMBC)}, 2020, pp. 3452--3455.

\bibitem{interrater}
H.~Danker-hopf \emph{et~al.}, ``Interrater reliability for sleep scoring
  according to the rechtschaffen \& kales and the new {AASM} standard,''
  \emph{J. Sleep Res.}, vol.~18, no.~1, pp. 74--84, 2009.

\bibitem{won2017motion}
D.-O. Won, H.-J. Hwang, D.-M. Kim, K.-R. M{\"u}ller, and S.-W. Lee,
  ``Motion-based rapid serial visual presentation for gaze-independent
  brain-computer interfaces,'' \emph{IEEE Trans. Neural Syst. Rehabilitation
  Eng.}, vol.~26, no.~2, pp. 334--343, 2017.

\bibitem{zhang2017hybrid}
Y.~Zhang, H.~Zhang, X.~Chen, S.-W. Lee, and D.~Shen, ``Hybrid high-order
  functional connectivity networks using resting-state functional {MRI} for
  mild cognitive impairment diagnosis,'' \emph{Sci. Rep.}, vol.~7, no.~1, pp.
  1--15, 2017.

\bibitem{thung2018conversion}
K.-H. Thung, P.-T. Yap, E.~Adeli, S.-W. Lee, D.~Shen, A.~D.~N. Initiative
  \emph{et~al.}, ``Conversion and time-to-conversion predictions of mild
  cognitive impairment using low-rank affinity pursuit denoising and matrix
  completion,'' \emph{Med. Image Anal.}, vol.~45, pp. 68--82, 2018.

\bibitem{lee2018high}
M.-H. Lee, J.~Williamson, D.-O. Won, S.~Fazli, and S.-W. Lee, ``A high
  performance spelling system based on {EEG}-{EOG} signals with visual
  feedback,'' \emph{IEEE Trans. Neural Syst. Rehabilitation Eng.}, vol.~26,
  no.~7, pp. 1443--1459, 2018.

\bibitem{lee2019connectivity}
M.~Lee, B.~Baird, O.~Gosseries, J.~O. Nieminen, M.~Boly, B.~R. Postle,
  G.~Tononi, and S.-W. Lee, ``Connectivity differences between consciousness
  and unconsciousness in non-rapid eye movement sleep: a {TMS}--{EEG} study,''
  \emph{Sci. Rep.}, vol.~9, no.~1, pp. 1--9, 2019.

\bibitem{kim2019subject}
K.-T. Kim, C.~Guan, and S.-W. Lee, ``A subject-transfer framework based on
  single-trial {EMG} analysis using convolutional neural networks,'' \emph{IEEE
  Trans. Neural Syst. Rehabilitation Eng.}, vol.~28, no.~1, pp. 94--103, 2019.

\bibitem{kwon2019subject}
O.-Y. Kwon, M.-H. Lee, C.~Guan, and S.-W. Lee, ``Subject-independent
  brain--computer interfaces based on deep convolutional neural networks,''
  \emph{IEEE Trans. Neural Netw. Learn. Syst.}, vol.~31, no.~10, pp.
  3839--3852, 2019.

\bibitem{jeong2020brain}
J.-H. Jeong, K.-H. Shim, D.-J. Kim, and S.-W. Lee, ``Brain-controlled robotic
  arm system based on multi-directional {CNN}-{B}i{LSTM} network using {EEG}
  signals,'' \emph{IEEE Trans. Neural Syst. Rehabilitation Eng.}, vol.~28,
  no.~5, pp. 1226--1238, 2020.

\bibitem{deepsleepnet}
A.~Supratak, H.~Dong, C.~Wu, and Y.~Guo, ``Deep{S}leep{N}et: A model for
  automatic sleep stage scoring based on raw single-channel {EEG},'' \emph{IEEE
  Trans. Neural Syst. Rehabil. Eng.}, vol.~25, no.~11, pp. 1998--2008, 2017.

\bibitem{attnsleep}
E.~Eldele \emph{et~al.}, ``An attention-based deep learning approach for sleep
  stage classification with single-channel {EEG},'' \emph{IEEE Trans. Neural
  Syst. Rehabil. Eng.}, vol.~29, no.~1, pp. 809--818, 2021.

\bibitem{xsleepnet}
H.~Phan \emph{et~al.}, ``{XS}leep{N}et: Multi-view sequential model for
  automatic sleep staging,'' \emph{IEEE Trans. Pattern Anal. Mach. Intell.
  (Early Access)}, 2021.

\bibitem{cesari2021interrater}
M.~Cesari \emph{et~al.}, ``Interrater sleep stage scoring reliability between
  manual scoring from two european sleep centers and automatic scoring
  performed by the artificial intelligence--based stanford-stages algorithm,''
  \emph{J. Clin. Sleep Med.}, vol.~17, no.~6, pp. 1237--1247, 2021.

\bibitem{zheng2016improving}
S.~Zheng, Y.~Song, T.~Leung, and I.~Goodfellow, ``Improving the robustness of
  deep neural networks via stability training,'' in \emph{Proc. IEEE Comput.
  Soc. Conf. Comput. Vis. Pattern Recognit.}, 2016, pp. 4480--4488.

\bibitem{transformer}
A.~Vaswani \emph{et~al.}, ``Attention is all you need,'' in \emph{Adv. Neural
  Inf. Process. Syst. (NIPS)}, 2017, pp. 5998--6008.

\bibitem{siam}
G.~Koch \emph{et~al.}, ``Siamese neural networks for one-shot image
  recognition,'' in \emph{ICML deep learning workshop}, vol.~2, no.~1, 2015.

\bibitem{skd}
S.~Yun, J.~Park, K.~Lee, and J.~Shin, ``Regularizing class-wise predictions via
  self-knowledge distillation,'' in \emph{Proc. IEEE Comput. Soc. Conf. Comput.
  Vis. Pattern Recognit. (CVPR)}, 2020, pp. 13\,876--13\,885.

\bibitem{bert}
J.~Devlin, M.-W. Chang, K.~Lee, and K.~Toutanova, ``Bert: Pre-training of deep
  bidirectional transformers for language understanding,'' \emph{arXiv preprint
  arXiv:1810.04805}, 2018.

\bibitem{vit}
D.~Alexey \emph{et~al.}, ``An image is worth 16x16 words: Transformers for
  image recognition at scale,'' \emph{CoRR}, vol. abs/2010.11929, no.~1, 2020.

\bibitem{dylee}
D.-Y. Lee, J.-H. Jeong, K.-H. Shim, and S.-W. Lee, ``Decoding movement
  imagination and execution from {EEG} signals using {BCI}-transfer learning
  method based on relation network,'' in \emph{Proc. IEEE Int. Conf. Acoust.
  Speech Signal Process. (ICASSP)}, 2020, pp. 1354--1358.

\bibitem{labelsmooting}
L.~Yuan, F.~E. Tay, G.~Li, T.~Wang, and J.~Feng, ``Revisiting knowledge
  distillation via label smoothing regularization,'' in \emph{Proc. IEEE
  Comput. Soc. Conf. Comput. Vis. Pattern Recognit. (CVPR)}, 2020, pp.
  3903--3911.

\bibitem{utime}
M.~Perslev, M.~H. Jensen, S.~Darkner, P.~J. Jennum, and C.~Igel, ``U-time: A
  fully convolutional network for time series segmentation applied to sleep
  staging,'' \emph{arXiv preprint arXiv:1910.11162}, 2019.

\bibitem{sleepedf}
B.~Kemp \emph{et~al.}, ``Analysis of a sleep-dependent neuronal feedback loop:
  the slow-wave microcontinuity of the {EEG},'' \emph{IEEE Trans. Biomed.
  Eng.}, vol.~47, no.~9, pp. 1185--1194, 2000.

\bibitem{shhs}
G.-Q. Zhang \emph{et~al.}, ``The national sleep research resource: towards a
  sleep data commons,'' \emph{J. Am. Med. Inf. Assoc.}, vol.~25, no.~10, pp.
  1351--1358, 2018.

\bibitem{adam}
D.~P. Kingma and J.~Ba, ``Adam: A method for stochastic optimization,''
  \emph{arXiv preprint arXiv:1412.6980}, 2014.

\bibitem{bhlee}
B.-H. Lee, J.-H. Jeong, K.-H. Shim, and S.-W. Lee, ``Classification of
  high-dimensional motor imagery tasks based on an end-to-end role assigned
  convolutional neural network,'' in \emph{Proc. IEEE Int. Conf. Acoust. Speech
  Signal Process. (ICASSP)}, 2020, pp. 1359--1363.

\bibitem{gelu}
D.~Hendrycks and K.~Gimpel, ``Gaussian error linear units {(GELUs)},''
  \emph{arXiv preprint arXiv:1606.08415}, 2016.

\bibitem{suk2014predicting}
H.-I. Suk, S.~Fazli, J.~Mehnert, K.-R. M{\"u}ller, and S.-W. Lee, ``Predicting
  {BCI} subject performance using probabilistic spatio-temporal filters,''
  \emph{PloS One}, vol.~9, no.~2, p. e87056, 2014.

\bibitem{jeong2020decoding}
J.-H. Jeong, N.-S. Kwak, C.~Guan, and S.-W. Lee, ``Decoding movement-related
  cortical potentials based on subject-dependent and section-wise spectral
  filtering,'' \emph{IEEE Trans. Neural Syst. Rehabilitation Eng.}, vol.~28,
  no.~3, pp. 687--698, 2020.

\end{thebibliography}

\end{document}